\documentclass[10pt,dvips]{article}
\usepackage{amssymb,amsfonts,amsmath,latexsym}
\usepackage{graphics,graphicx,epsf}
\newcommand{\be}{\begin{equation}}
\newcommand{\ee}{\end{equation}}
\newcommand{\bea}{\begin{eqnarray}}
\newcommand{\eea}{\end{eqnarray}}
\newcommand{\ba}{\begin{array}}
\newcommand{\ea}{\end{array}}
\newcommand{\bt}{\begin{tabular}}
\newcommand{\et}{\end{tabular}}

\newcommand{\fr}{\frac}
\newcommand{\ci}{\cite}
\newcommand{\cl}{\centerline}
\newcommand{\bs}{\bigskip}

\newcommand{\vs}{\vspace}

\newcommand{\en}{\eqno}

\newcommand{\fns}{\footnotesize}

\newcommand{\bbib}{\begin{thebibliography}}
\newcommand{\ebib}{\end{thebibliography}}

\newcommand{\und}{\underline}

\begin{document}
\bs
\cl{\bf CLASSICAL HALL TRANSITION AND MAGNETORESISTANCE}
\cl{\bf IN STRONGLY INHOMOGENEOUS PLANAR SYSTEMS}

\bs

\cl{\bf S.A.Bulgadaev \footnote{e-mail: bulgad@itp.ac.ru}
}
\bs \cl{\fns Landau Institute for Theoretical Physics,
Chernogolovka, Moscow Region, Russia, 142432}
%\cl{\fns
%Department of Physics, Loughborough University, Loughborough, LE11
%3TU, UK}

\bs

\begin{quote}
\footnotesize{ The magneto-transport properties of 
planar and layered strongly inhomogeneous two-phase systems are
investigated, using the explicit expressions for the effective conductivities and resistivities obtained by the exact dual transformation, connecting effective conductivities of in-plane isotropic two-phase systems with and
without magnetic field. These expressions allow to describe the effective resistivity of various inhomogeneous media at arbitrary
concentrations $x$ and magnetic fields $H$.  The corresponding
plots of the $x$-dependence of the Hall constant $R_H(x,H)$ and the magnetoresistance $R(x,H)$ are constructed for various values of magnetic field at some values of inhomogeneity parameters. 
These plots for strongly inhomogeneous systems at high magnetic fields show a sharp  transition between partial Hall resistivities (or Hall conductivities) with different dependencies of $R_H$ on the phase concentrations.
It is shown that there is a strong correlation between large
linear magnetoresistance effect and this sharp Hall transition. 
Both these effects are a consequence of the exact duality symmetry.
A possible physical explanation of these effects and their correlation is proposed.}
\end{quote}

\bs
\cl{PACS: 75.70.Ak, 72.80.Ng, 72.80.Tm, 73.61.-r}
\bs

\underline{\bf 1. Introduction}

\bs

The experimental discovery of the abnormal magnetotransport
properites in thin films of some inhomogeneous systems even at room
temperatures $T$ and high magnetic fields $H,$ such as a large \ci{1}
and a linear magnetoresistance (MR) \ci{2}, has turned out a challenge for theoretical
physics, since the existing theories of the magnetotransport in
inhomogeneous systems (see, for example \ci{3}) (using presumably the effective medium approximation, based on the wire-network type model of
inhomogeneous media \ci{4}) cannot describe properly these
properties \ci{3,5}.

Recently, the new network type model, generalizing the standart wire-network
type model and using as the building blocks the 4-terminal discs,
was proposed \ci{5,6}. This model can describe magnetotransport
properties of thin films. The authors of this model have shown 
by numerical methods that the disc-network model can
reproduce the large linear MR (LLMR), when the conducting parameters of the discs, in particular, a mobility $\mu,$ have random continuous and wide
distribution with an average value $\langle \mu \rangle = 0.$ It
was noted also in \ci{5,6}, that the investigation of the Hall
resistivity of this model by the same methods require much more
time and work.

In the framework of the classical theory  there is another
approach for a description of the magnetotransport properties of some
inhomogeneous two-dimensional (or layered, inhomogeneous in
planes, but constant in the direction orthogonal to planes)
systems in perpendicular magnetic field. It is based on the exact
dual symmetry of these systems \ci{7} and is
connected with an existence of the exact dual, linear fractional
transformation, relating the effective conductivity $\hat
\sigma_e$ (and the effective resistivity $\hat \rho_e = \hat
\sigma_e^{-1}$) of planar self-dual two-phase systems with and without ($\sigma_{e0}$)
magnetic field at arbitrary magnetic fields and phase
concentrations \ci{8,9}.

In the papers \ci{10,11,12}, using this exact transformation (we call
it the magnetic dual transformation (MDT)) and the known
expressions for $\sigma_e$ of three inhomogeneous models with
different random structures of inhomogeneities without $H$
from \ci{13}, the explicit expressions for $\hat \sigma_e$ \ci{10,11}
and for the magnetoresistance $R(x,H)$ \ci{12} have been obtained.
It have been shown also there (\ci{11,12}) that these expressions give the LLMR
effect even in two-phase systems and without the restriction
$\langle \mu \rangle =0,$ and that a form and a position of the LLMR
effect as  functions of phase concentrations depend on models and the corresponding inhomogeneity structures.

In this letter, using the explicit approximate expressions
obtained in \ci{10,11,12} and applicable at arbitrary values of phase
concentrations and magnetic fields and in a wide region of partial
conductivities, we will investigate a behaviour of the effective resistivities $\rho_{et}$ and $\rho_{et}$ of the planar self-dual
two-phase systems. We will present also the $x$-dependence
plots of the Hall constant $R_H(x,H)$ for some characteristic values
of magnetic field $H$ at different values of the inhomogeneity parameters. 
These plots, together with the previous results for the effective
conductivity from \ci{10,11} and for the magnetoresistance $R(x,H)$
from \ci{12}, show the existence of the following properties in
these inhomogeneous systems at high magnetic fields: 1) the sharp
transition between partial Hall constant plateaus under a change of phase
concentrations with a position and a width of the region of
transition depending on the model inhomogeneity structure; 2) the
strong correlations between the LLMR effect and the sharp
transition between the Hall plateaus. These properties are a consequence of 
the exact duality relation. A possible
physical explanation of these properties is also proposed. A
comparison with the existing experimental data shows a good
qualitative agreement.

\bs

\und{\bf 2. Magnetic dual transformation at high $H$}

\bs

The effective conductivity of inhomogeneous isotropic systems in
magnetic field has the following form
$$
\hat \sigma_e = \sigma_{e(ik)} = \sigma_{ed} \delta_{ik} +
\sigma_{et} \epsilon_{ik}, \quad \sigma_{ed} ({\bf H}) =
\sigma_{ed} (-{\bf H}), \quad \sigma_{et} ({\bf H}) = -\sigma_{et}
(-{\bf H}), \en(1)
$$
here $\delta_{ik}$ is the Kronecker symbol, $\epsilon_{ik}$ is the unit antisymmetric tensor.
The effective resistivity $\hat \rho_{e}$ is defined by the inverse matrix
$$
\hat \rho_e = \rho_{e(ik)} = \rho_{ed} \delta_{ik} + \rho_{et}
\epsilon_{ik}, \quad \rho_{ed} ({\bf H}) = \rho_{ed} (-{\bf H}),
\quad \rho_{et} ({\bf H}) = -\rho_{et} (-{\bf H}), \en(2)
$$
where
$$
\rho_{ed} = \sigma_{ed}/(\sigma_{ed}^2 + \sigma_{et}^2), \quad
\rho_{et} = -\sigma_{et}/(\sigma_{ed}^2 + \sigma_{et}^2). 
\en(3)
$$
The Hall constant $R_H(x,H)$ is determined as follows
$$
R_H(x,H) = \rho_{et}(x,H)/ H. \en(4)
$$
In two-phase systems the effective conductivity $\hat \sigma_e$ (and
effective resistivity $\hat \rho_e$) can be expressed by the
"magnetic" dual transformation (MDT) through the  effective
conductivity of the same system without magnetic field \ci{8,9}
$$
\sigma_{ed} (\{\sigma\}, \{x\}) =
\fr{\sigma'_{ed}(ac+b)}{(\sigma'_{ed})^2 + a^2}, \quad
\sigma_{et} (\{\sigma\}, \{x\}) = c \fr{(\sigma'_{ed})^2 -a
b'}{(\sigma'_{ed})^2 + a^2},
\en(5)
$$
where $\sigma'_{ed}$ is the effective conductivity of the models
without magnetic field, but it depends on transformed partial
arguments $\sigma'_{id} = \sigma_{id}/\sigma_{ai}$ with
$\sigma_{ai} = \sigma_{it} + a.$ The parameters of the MDT
$a,b'=b/c,c$ depend on the partial conductivities and have the
following form
$$
a_{\pm} = \fr{|\sigma_2|^2 - |\sigma_1|^2 \pm
\sqrt{B}}{2(\sigma_{1t} - \sigma_{2t})}, \quad b'_{\pm} =
\fr{|\sigma_1|^2 - |\sigma_2|^2 \pm \sqrt{B}}{2(\sigma_{1t} -
\sigma_{2t})},\quad c=-a,
$$
$$
B = [(\sigma_{1t} - \sigma_{2t})^2 + (\sigma_{1d} -
\sigma_{2d})^2] [(\sigma_{1t} - \sigma_{2t})^2 + (\sigma_{1d} +
\sigma_{2d})^2],
\en(6)
$$
where $|\sigma_i|^2 = \sigma_{id}^2 + \sigma_{it}^2,$ and,
evidently, $B\ge 0.$ Note that $H$ enters in (,) only through the
partial conductivities $\hat \sigma_i$ and $\sigma_{ai}$

The effective conductivities $\sigma_{ed}$ and  $\sigma_{et}$
obtained by the MDT belongs to the
circle  in the $\sigma$-plane with a radius $R$ and centered at the point $(0,z_0)$ \ci{8,9} (called a "semicircle law " in \ci{14})
$$
\sigma_{ed}^2 + (\sigma_{et} - z_0)^2 = R^2,
\en(7)
$$
where
$$
z_0 = \fr{|\sigma_1|^2 - |\sigma_2|^2}{2(\sigma_{1t} -
\sigma_{2t})} , \quad R = \fr{\sqrt{B}}{2(|\sigma_{1t} -
\sigma_{2t}|)}.
\en(8)
$$
Here $z_0$ can be positive as well as negative in dependence of values of
partial conductivities. The circle (7) does not contain $x$ and $H$ explicitly
and goes through the partial conductivities $\hat \sigma_i, \; (i=1,2)$.
The arc constrained by the points $\hat \sigma_i, \; (i=1,2)$ can be named as the effective conductivity arc, since $\hat \sigma_e$ maps the concentration segment $0 \le x \le 1$ into it (see Figure 1a). In these formulas one assumes that $\sigma_{1t} \ne \sigma_{2t},$
since for $\sigma_{1t} = \sigma_{2t}$ system effectively reduces to that without magnetic field \ci{15}. In terms of the dual transformations $T$ \ci{7,8,9} this case corresponds to a shift of $\sigma_{e0}$ on $\sigma_{1t}$ along $\sigma_t$ axis
$$
T_{sh} \sigma_{e0} (\sigma_i) = \sigma_{e0} \left( T_{sh} \sigma_i \right), \quad
T_{sh} \sigma_i = T_{sh} \sigma_{id}\delta_{ik} = \sigma_{id}\delta_{ik} + \sigma_{1t}\epsilon_{ik}.
$$
Below, for a definiteness, we assume that $\sigma_{id}, \sigma_{1t} \ge 0$ and $\sigma_{1d} \ge \sigma_{2d},\; \sigma_{1t} \ge \sigma_{2t}$. As one can see from the Fig.1a, the properties of $\hat \sigma_e$ 
will depend essentially on a position of the arc on the circle (7), which is determined by the partial conductivities $\hat \sigma_i \; (i=1,2).$ Due to (7),
$\sigma_{et}$ must be a monotonous function of $x,$ changing from $\sigma_{2t}$
at $x=0$ up to $\sigma_{1t}$ at $x=1.$ At the same time, in contrast with a case $H=0$, in the inhomogeneous systems with $H \ne 0$ there is a possibility for $\sigma_{ed}$ to be larger than the largest partial conductivity $\sigma_{1d},$
and a dependence of $\sigma_{ed}$ on $x$ can be non-monotonous with a maximum at some $0 \le x_m \le 1.$ It follows from (7) that the maximal value $\sigma_{ed}$ can take is
$$
\sigma_{ed}(x_m) = \sigma_{dm} = R.
\en(9)
$$
It can be realized only when $z_0$ is between the boundary points of the arc, i.e.  $\sigma_{1t} \ge z_0 \ge \sigma_{2t}$ (or in the opposite order if $\sigma_{1t} \le \sigma_{2t}$). At $x_m$ the Hall conductivity $\sigma_{et}(x_m) = z_0.$ Both these
values are exact, but a position of $x_m$ depends on a model.
\begin{figure}[t]
\begin{picture}(250,240)
\put(50,0){%
\begin{picture}(80,80)
\put(10,0){\vector(0,1){100}}
\put(10,10){\vector(1,0){60}}
\put(10,105){$\sigma_{t}$}
\put(75,10){$\sigma_{d}$}
\qbezier(10,10)(58,12)(60,50)
\qbezier(10,90)(58,88)(60,50)
\put(-10,50){$z_0$}
\put(10,50){\circle*{3}}
\put(27,12){\circle*{3}}
\put(56,33){\circle*{3}}
\put(20,22){$\hat \sigma_{2}$}
\put(66,33){$\hat \sigma_{1}$}
\put(32,-18){({\small a})}
\end{picture}}

\put(210,0){%
\begin{picture}(80,80)
\put(10,0){\vector(0,1){100}}
\put(10,10){\vector(1,0){60}}
\put(10,105){$\sigma_{t}$}
\put(75,10){$\sigma_{d}$}
\qbezier(10,12)(58,14)(60,52)
\qbezier(10,92)(58,90)(60,52)
\put(-10,52){$z_0$}
\put(10,52){\circle*{3}}
\put(12,12){\circle*{3}}
\put(12,92){\circle*{3}}
\put(-10,12){$\sigma_{2t}$}
\put(-10,92){$\sigma_{1t}$}
\put(32,-18){({\small b})}
\end{picture}}

\end{picture}

\vs{1cm}

{\small Fig.1. (a) A schematic picture of the right part of the circle (7) in the $\sigma_{d,t}$-plane, when $\sigma_{1d,t} \ge \sigma_{2d,t} \ge 0.$ (b) The same semicircle (rescaled and magnified) in the strong magnetic field limit.
}
\end{figure}
For systems with the opposite signs of $\sigma_{it}$ (i.e. with the opposite signs of charge carriers in partial components)
there exists always a concentration $x_0$ at which $\sigma_{et}$ changes a sign, i.e. $\sigma_{et}(x_0) = 0.$  The corresponding value $\sigma_{ed}(x_0)$ 
$$
\sigma_{ed}(x_0) = \sigma_{d0} = \sqrt{R^2 - z_0^2}
\en(10)
$$
is also an exact and does not exceed  $\sigma_{dm},$ coinciding with the latter only when $z_0 =0.$

The effective resistivities also satisfy an analogous "semicircle law" 
$$
\rho_{ed}^2 + (\rho_{et} - z_{\rho})^2 = R_{\rho}^2,
\en(11)
$$
where
$$
z_{\rho} = \fr{z_0}{R^2 - z_0^2}, \quad R_{\rho} = \fr{R}{|R^2 -z_0^2|}.
\en(12)
$$
For the effective Hall constant $R_H$ this relation can be written as a "straight line law" \ci{16,8}
$$
\fr{R_{H2} - R_{H}}{R_{H2} - R_{H1}} = \fr{\rho_{2d} - \rho_{ed} + H^2(R_{H2}^2 - R_{H}^2)}{\rho_{2d} - \rho_{1d} + H^2(R_{H2}^2 - R_{H1}^2)}.
\en(11')
$$
Consequently, $\rho_{ed,et}$ have the similar properties: a monotonous dependence of $\rho_{et}$ (and $R_H$) on $x$ and a possible non-monotonous behaviour of $\rho_{ed}$ with the exact maximal value at some $x_{m1}$
$$
\rho_{ed}(x_{m1}) = \rho_{dm} = R_{\rho}.
\en(13)
$$
Again, $\rho_{et}$ and $R_{H}$ will change a sign at some $x_{01}$ for systems with $R_{H1}R_{H2} \le 0,$ where $\rho_{ed}(x_{01})= \rho_{d0} = \sqrt{R^2_{\rho} - z^2_{\rho}}.$

In order to describe the dependence of the $\hat \sigma_{e}$ and $\hat \rho_{e}$ of inhomogeneous systems on magnetic field one needs to know the
dependences  of partial conductivities on $H$
$\sigma_{id}({\bf H}), \sigma_{it}({\bf H}) \; (i=1,2).$ The partial conductivities are approximated usually by the standard (metallic type) formulae \ci{7}
$$
\sigma_{id}({\bf H}) = \fr{\sigma_{i0}}{1+ \beta_i^2}, \quad
\sigma_{it}({\bf H}) = \fr{\sigma_{i0} \beta_i}{1+ \beta_i^2}, \quad
\beta_i = \mu_{i} H, \quad i=1,2,
\en(14)
$$
where $\sigma_{i0}$ are the partial conductivities of phases
without magnetic field, $\mu_i = e_i \tau_i/m_i$ are the corresponding mobilities,
$\tau_i, e_i, m_i$ are, respectively,  carrier's times of life, charges and masses.
Analogous formulas for $\rho_i$ and $R_{Hi}$ are
$$
\rho_{id}({\bf H}) = \fr{1}{\sigma_{i0}}, \quad
\rho_{it}({\bf H}) = R_{Hi} H, \quad
R_{Hi} = -\mu_i/\sigma_{i0} = -1/e_in_i , \quad i=1,2.
\en(15)
$$ 
Note that the formulas (14,15) have already an asymptotic form corresponding to high $H$ (since $\rho_{id}$ do not depend on $H,$  $\rho_{it}$ are linear in $H,$ and $R_{Hi}$ depend only on the carrier's charges and concentrations $n_i$).
They defines the boundary values of $\hat \sigma_e, \;\hat \rho_e,$ and $R_H(x,H)$ at $x=0,1$  for inhomogeneous systems, satisfying the representation (14). Since the mobilities enter always in a combination with a magnetic field,
 $\hat \sigma_{i}$ depends really
(besides a renormalized magnetic field $H' = \mu_1 H$, below we will suppose $\mu_1 = 1,$ because it defines only a scale of $H$ where a crossover from effectively low to high $H$ for the phase one takes place)
on two inhomogeneity parameters $z$  and $\eta$
(we suppose also that $\sigma_{20}/\sigma_{10} \le 1$)
$$
z = \sigma_{20}/\sigma_{10} \quad (0 \le z \le 1), \quad
\eta = \mu_2/\mu_1 \quad (-\infty \le \eta \le \infty).
\en(16)
$$
Note that the restriction  $\langle \mu \rangle =0$ corresponds to the case 
$\eta = -1.$ 

In a general case all parameters and $H$ have the same
order $\sim 1.$ Since below we will be interested presumably by a
strong magnetic fields limit, we need to know general properties
of the MDT and $\hat \sigma_e, \; \hat \rho_e$ at high $H.$ These properties
become more simple and evident in the limit
$$
H, \eta H \ge 1,
\en(17)
$$
when one has for partial conductivities
$$
\sigma_{1d} \sim 1/H^2, \quad \sigma_{2d} \sim z/\eta^2 H^2, \quad
\sigma_{1t} \sim 1/H, \quad \sigma_{2t} \sim z/\eta H,
\en(18)
$$
It follows from (18) that in this limit $\sigma_{1t} \gg \sigma_{1d}$ and $\sigma_{2t} \gg \sigma_{2d}.$
Further we will confine ourselves by the limit (18) and a case, when
both partial Hall parts $\sigma_{it}$ are much larger than all 
diagonal parts ($\sigma_{1,2t} \gg \sigma_{1,2d}$). In this case $\sigma_{it}$ will determine the magneto-transport properties of the inhomogeneous
systems. Nevertheless, even in this limit there are different cases connected with relations between the inhomogeneity parameters: (i) $z/\eta \sim 1$, (ii) $z/\eta \ll 1$, (iii) $z/\eta \gg 1$. Below we will consider the cases (i) and (ii) only.

At high $H$ the parameters of the MDT take, for
example in the case (i), have the following simple form
$$
a = -\sigma_{2t} + \fr{\sigma_{2d}^2}{\sigma_{t-}}, \quad
b'= -\sigma_{1t} + \fr{\sigma_{1d}^2}{\sigma_{t-}}, \quad
$$
$$
z_0=\sigma_{t+}/2 + \fr{\sigma_{1d}^2 - \sigma_{2d}^2}{\sigma_{t-}}, \quad
R=\fr{|\sigma_{t-}|}{2} \left(1+\fr{\sigma_{1d}^2 + \sigma_{2d}^2}{\sigma_{t-}^2}\right),
$$
$$
\sigma_{a1}= \sigma_{t-} + \fr{\sigma_{2d}^2}{\sigma_{t-}}, \quad
\sigma_{a2} = \fr{\sigma_{2d}^2}{\sigma_{t-}},
\quad \sigma_{t\pm} = \sigma_{1t} \pm \sigma_{2t}.
\en(19)
$$
Analogous formulas take place in this limit for $z_{\rho}$ and $R_{\rho}$
with the corresponding substitutions of $\rho_{it}$ instead of $\sigma_{it}$ (see
the sections 3,4).
One can see from (19) that in the high $H$ limit $z_0$ is always situated between $\sigma_{it}.$ This fact allows a realization of the non-monotonous $x$-dependence of $\sigma_{ed}$ with a maximal
value (9) at some intermediate concentration $x_m.$ Moreover, though $\sigma_{it} \gg \sigma_{id}$ at high
magnetic fields, $\sigma_{ed}(x_m)= R \approx |\sigma_{1t} - \sigma_{2t}|/2$ has the same order as $\sigma_{it}$ (when $\sigma_{it}$ are of the same order) and the order of a maximal
$\sigma_{1t}$ (when $\sigma_{1t} \gg \sigma_{2t}$). Thus, due to the exact "semicircle" condition (7), there is always at high $H$ a region of concentrations, where
$\sigma_{dm} \sim \sigma_{it} \sim H^{-1}.$ A position and a
width of this region depend on a model and on the structure of its inhomogeneities \ci{10,11}. One such example is shown in \ci{10} on fig.2,
when $\sigma_{it}$ have the opposite signs and a maximum of $\sigma_{ed}$ is
situated near $x=1.$ For this reason, one cannot neglect $\sigma_{ed}$ in such inhomogeneous systems even if their $\sigma_{it} \gg \sigma_{id}!$
Just this fact will give a solution of the long-standing puzzle of the LLMR effect.

Since the exact values for effective conductivity  at equal phase concentrations $x=1/2$ are known (see, for example, \ci{9})
$$
\sigma_{ed}(x=1/2)=\sqrt{\sigma_{1d} \sigma_{2d}} A, \quad
A = \left[1+\left(\fr{\sigma_{1t} - \sigma_{2t}}{\sigma_{1d} + \sigma_{2d}}\right)^2 \right]^{1/2},
$$
$$
\sigma_{et}(x=1/2)=\fr{\sigma_{2t} \sigma_{1d} + \sigma_{1t} \sigma_{2d}}{ \sigma_{1d} + \sigma_{2d}},
\en(20)
$$
one can write the exact expression for $\rho_{et}$ and $R_H$ at $x=1/2.$
A direct usage of (20) gives rather complicate expressions, for this reason we will use an alternative, more simple, exact expression for $\rho_{et}(1/2,H),$ which follows from the fact, that $\hat \sigma_{e}(1/2,H)$ is a fixed point of the DT, interchanging phases \ci{9},
$$
\rho_{ed}(1/2,H) = \fr{\sqrt{-{a^*}^2 +b^*}}{b^*} = \fr{\sigma_{ed}(1/2,H)}{b^*}, \quad
\rho_{et}(1/2,H) = \fr{a^*}{b^*},
$$
$$
a^* = - \sigma_{et}(1/2,H), \quad b^* = \sigma_{ed}^2(1/2,H) +\sigma_{et}^2(1/2,H)=
$$
$$
= \fr{\sigma_{1d}\sigma_{2t}^2 + \sigma_{2d}\sigma_{1t}^2 + \sigma_{1d}\sigma_{2d} (\sigma_{1d} + \sigma_{2d})}{\sigma_{1d} + \sigma_{2d}}
\en(21)
$$
The asymptotic behaviour of $\rho_{et}$ at $H, \eta H \gg 1$ is
$$
\rho_{et}(1/2,H)=  R_H(1/2,H) H + o(H), \quad
R_H(1/2,H) =\sigma_{10}^{-1} \fr{1+\eta}{1+z}.
\en(22)
$$
The fact that the exact value $\rho_{et}(1/2,H)$ has a linear dependence on $H$ at high $H$ was known \ci{7}. The Hall coefficient $R_H(1/2,H) =0$ only for inhomogeneous systems with $\eta = -1$ (this value corresponds in our case to the condition $\langle \mu \rangle =0$), when $\sigma_{et}(1/2,H) = \rho_{et}(1/2,H) = 0,$ while  $R_H(1/2,H)=1$ only at $\eta = z.$ 
\bs

\und{\bf 3. Sharp transition between Hall plateaus at high $H$.}

\bs

For a consideration of magneto-transport properties of strongly inhomogeneous planar systems we will use three explicit approximate expressions for the effective conductivity $\hat \sigma_e$ of planar two-phase systems
with different structures of inhomogeneities  obtained by the exact MDT in
\ci{9,10,11}. Two of them describe systems with various real "bulk"
inhomogeneities (i.e. having real 2D bulk inhomogeneities): the
compact inclusions  of regions of different sizes and forms
 of one phase into another (the "random droplets" model (RDM) and the "random parquet" structure constructed from square plaquettes with randomly
distributed stripes of two phases (the "random parquet" model (RPM))\ci{13}.  The third one has the effective conductivity,
obtained by the "magnetic" dual transformation from the known effective medium
approximation (EMA) formula, based on the wire-network
representation of inhomogeneous systems \ci{6} (we will name it the effective medium model (EMM)).

The explicit expressions of the effective conductivities $\sigma_{ed}$ and $\sigma_{et}$ of these inhomogeneous models are represented in \ci{10,11,12}.
All these formulas correctly reproduce boundary values of the
effective conductivities as well as their exact values at equal
phase concentrations $x=1/2$.

Substituting them into (3,4) one can
obtain the explicit expression for the Hall resistivity $\rho_{et}(x,H)$ and the Hall constant $R(x,H)$
of the corresponding systems at arbitrary concentrations and magnetic fields. Since these formulae are
rather complicate we will analyze their behaviour by constructing
the plots of their dependencies on $x$ and $H$.
The corresponding plots of $x$- and $H$-dependencies at $z=10^{-2}$ and $\eta = 1$ are represented, respectively, in Fig.2(a)-(c) and Fig.2(d).

\begin{figure}[t]
\begin{tabular}{cc}
\input epsf \epsfxsize=5.5cm \epsfbox{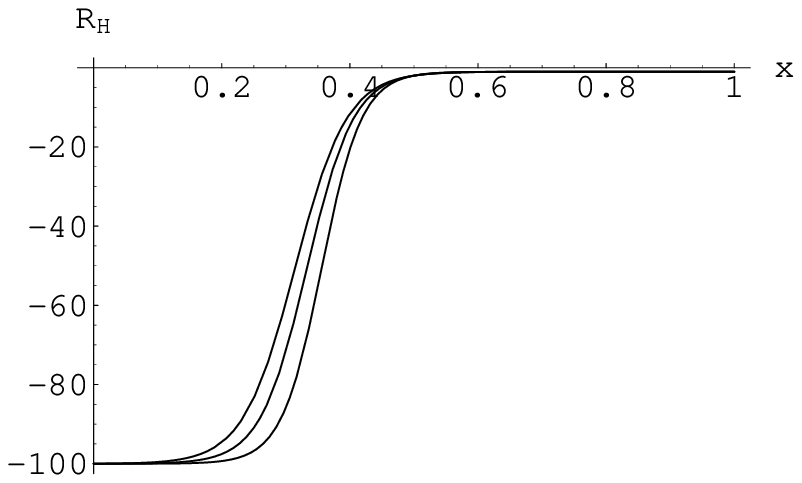}&
\input epsf \epsfxsize=5.5cm \epsfbox{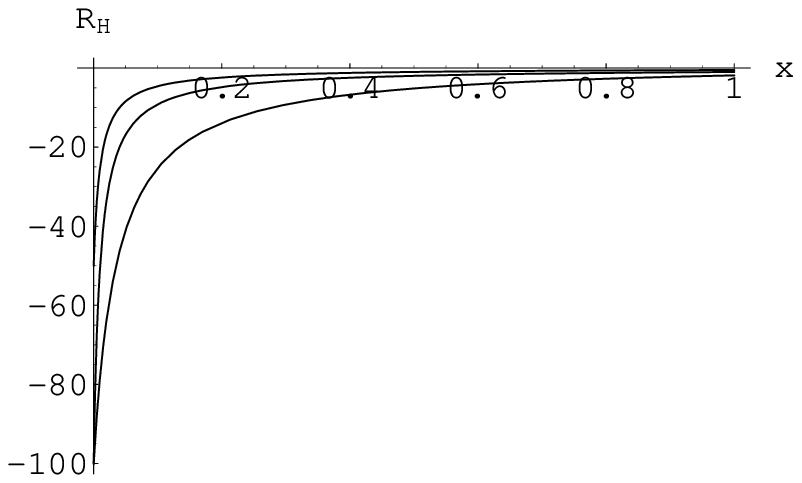}\\
{} & {}\\
(a) & (b)\\
\end{tabular}

\vs{0.3cm}

\begin{tabular}{cc}
\input epsf \epsfxsize=5.5cm \epsfbox{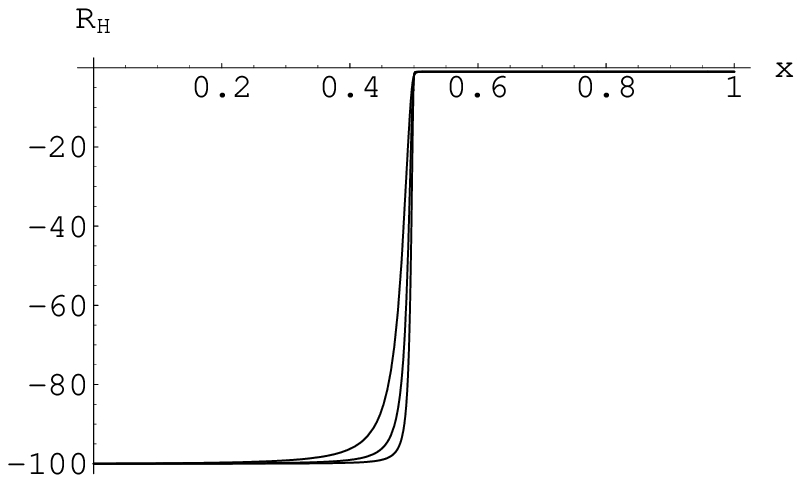}&
\input epsf \epsfxsize=5.5cm \epsfbox{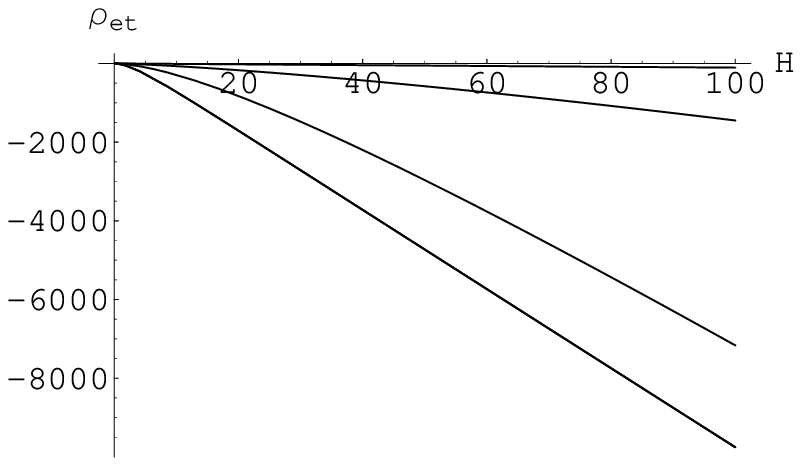}\\
{}&{}\\
(c)& (d)\\
\end{tabular}

\vs{0.3cm}

{\small  Fig.2.(a),(b),(c)  The  plots of the $x$-dependence of the Hall constant $R_H$ for three models at the inhomogeneity parameters
$z=0.01, \eta=1$ and at three different
(dimensionless) values of magnetic field $H$ (respectively,
(a),(b),(c)): 1) 50, 2) 100, 3) 300 (the corresponding plots go
from the left to the right for the "random droplet" and EM models and in the opposite direction for the "random parquet" model).
(d) shows, for a example, the $H$-dependence of $\rho_{et}$ for a "random droplets"  model at $z=0.01$ and some concentrations.}
\end{figure}
One can see from Fig.2a,c that the behaviour of $R_H$, though different for various models, has two common characteristic features for the RDM and EMM:

1) absolute values of $R_H$ at small $x$ and $1-x$ remain almost constant, with large and small values respectively, then rapidly changes
from large to small values in some narrow region of
concentrations, which width decreases with an increase of $H$; 

2) all values of  $R_H$ in the regions with a rapid change  slightly depend on $H,$ denoting a deviation from a pure linear dependence at intermediate $x.$

\begin{figure}[t]
\begin{tabular}{cc}
\input epsf \epsfxsize=5.5cm \epsfbox{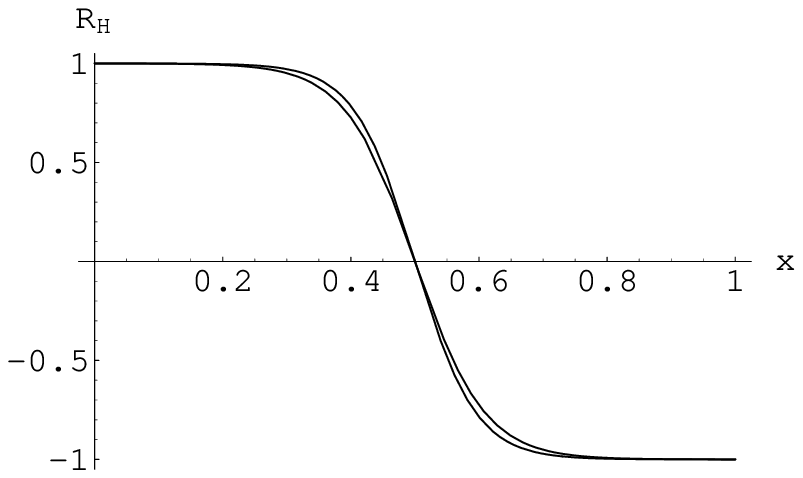}&
\input epsf \epsfxsize=5.5cm \epsfbox{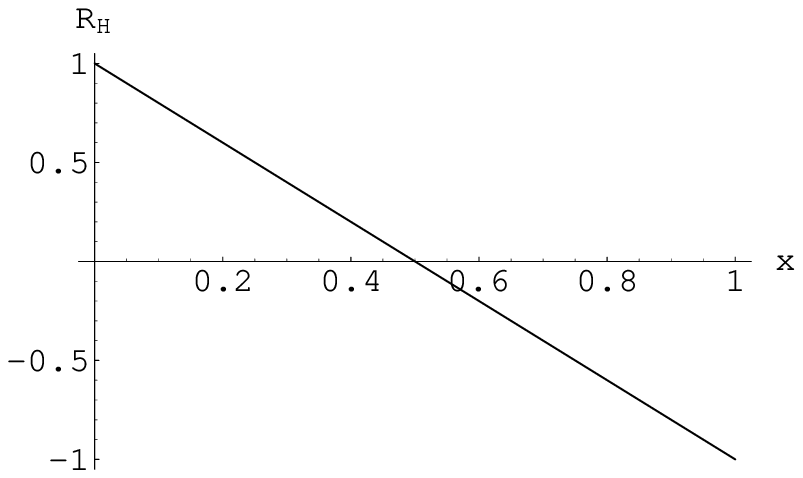}\\
{} & {}\\
(a) & (b)\\
\end{tabular}

\vs{0.3cm}

\begin{tabular}{cc}
\input epsf \epsfxsize=5.5cm \epsfbox{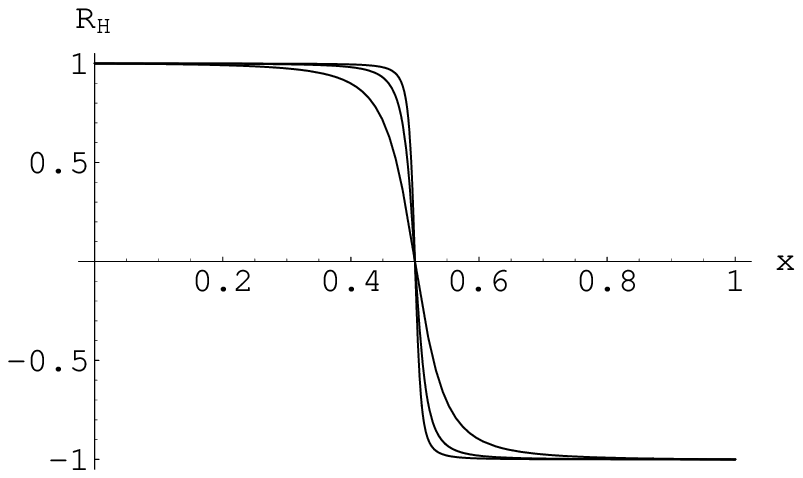}&
\input epsf \epsfxsize=5.5cm \epsfbox{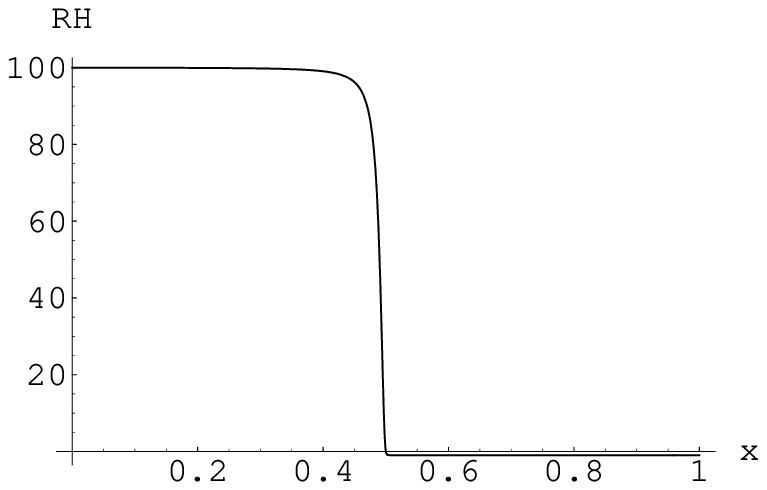}\\
{} & {}\\
(c) & (d)\\
\end{tabular}
\vs{0.3cm}
%\cl{(c)}
%\vs{0.3cm}

{\small  Fig.3 (a),(b),(c).  The  plots of the $x$-dependence of the
Hall constant $R_H(x,H)$ for three models (respectively, (a),(b),(c)) at the inhomogeneity parameters $z=1$ and $\eta = -1$ and at different
(dimensionless) values of magnetic field $H:$ 25, 50 for the RDM and RPM and 10, 25, 50 for EMM (the corresponding plots go from the left to the
right). (d)  shows for a comparison the plot of $R_H(x,H)$ at $z=0.01$ and $\eta=-1$ for EM model at $H=50.$}
\end{figure}

We will call such  behaviour the sharp Hall transition (SHT).
Practically the same plots  for $R_H(x,H)$ (except the overall sign) are obtained in the concentration region $0 \le x \le 0.5$ at $\eta=-1$ (see, for example, Fig.3d). The difference in the
region $0.5 \le x \le 1$ is relatively small and is connected with a change of a sign at $x = 1/2.$ This smalleness is connected with relatively small values of $R_H$ in the region $x \ge 1/2.$
 
We have checked also that $\rho_{ed}$ and $R_H$  satisfy  the "straight line" equation (11') for all three models with a very high accuracy. 
It is not strange because any map of the concentration segment into the effective conductivity arc must satisfy this equation.
\begin{figure}[t]
\begin{tabular}{cc}
\input epsf \epsfxsize=5.5cm \epsfbox{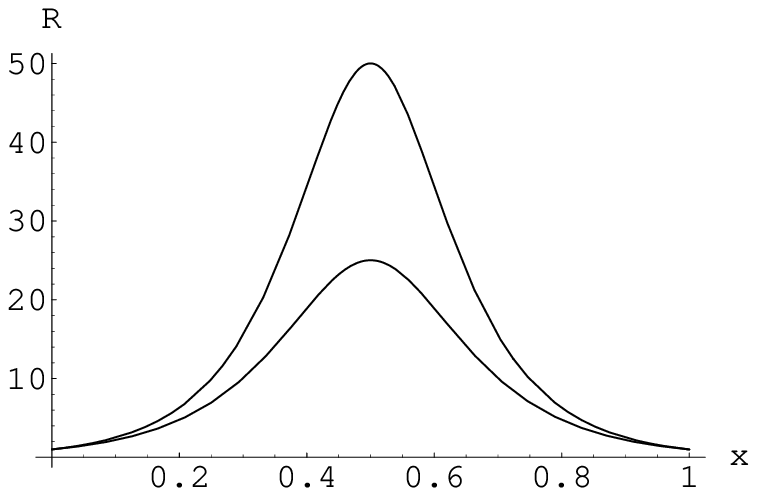}&
\input epsf \epsfxsize=5.5cm \epsfbox{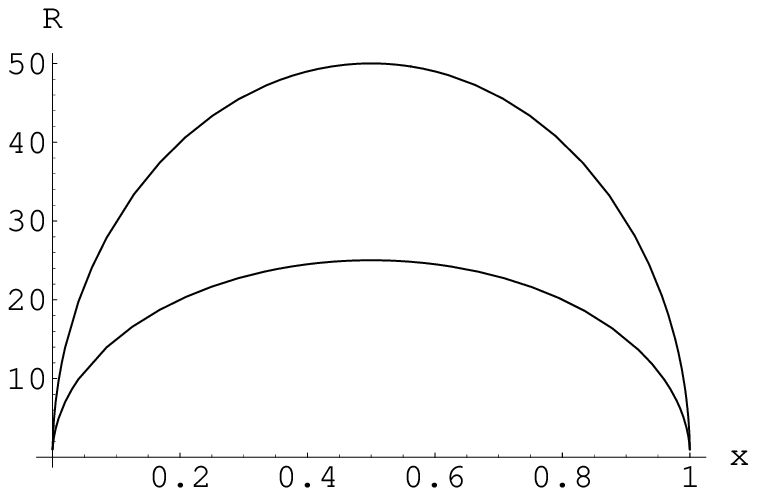}\\
{} & {}\\
(a) & (b)\\
\end{tabular}

\vs{0.3cm}

\begin{tabular}{cc}
\input epsf \epsfxsize=5.5cm \epsfbox{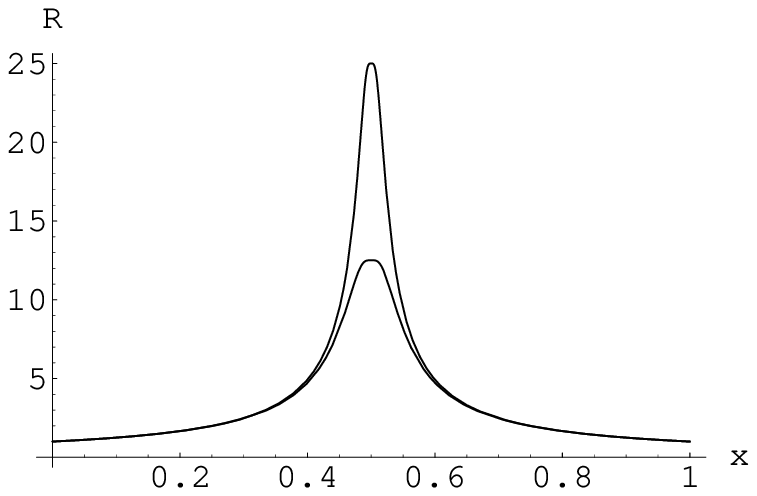}&
\input epsf \epsfxsize=5.5cm \epsfbox{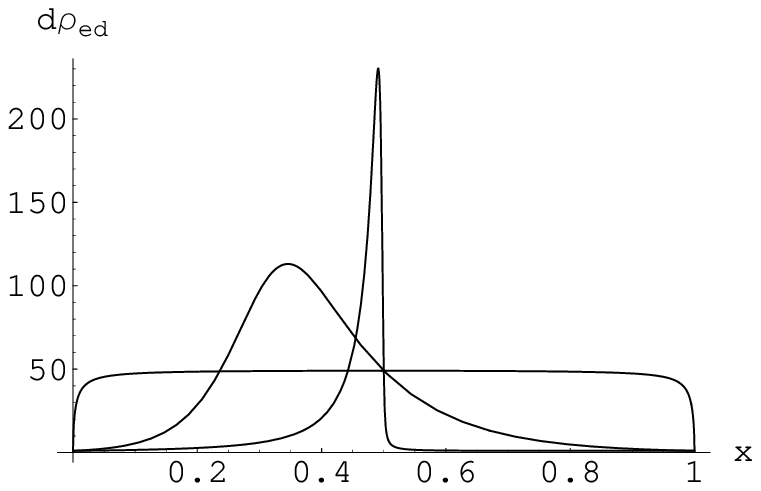}\\
{}&{}\\
(c)& (d)\\
\end{tabular}

\vs{0.3cm}

{\small  Fig.4.(a),(b),(c).  The  plots of the $x$-dependence of the
magnetoresistance $R(x,H)$ for three models (respectively, (a),(b),(c)) at the inhomogeneity parameters $z=1$ and $\eta = -1$ and at the three different
(dimensionless) values of magnetic field $H:$ 10,25 for the EMM and 25, 50 for the RDM and RPM (the corresponding plots go from the lower to the
upper ones).  
(d) shows for a comparison the plots of $R(x,H)$ at $z=0.01$ and $\eta=1$
for all models at $H=50.$}
\end{figure}

More detailed analysis shows a number of interesting specific properties. 
Let us start with the "random droplets" model. In this case $R_H$ has evident rather sharp  transition positioned for the inhomogeneity parameter $z=10^{-2}$ in the region of concentration $0.2 \le x \le 0.4,$ where the LLMR effect takes place, with a width of the transition region slightly  narrowing with an increase of $H$ and a decrease of inhomogeneity parameter $z$ (Fig.2a).  

The most striking plots have been obtained for the effective medium model (Fig.2c).
$R_H(x,H)$ shows two plateaus, corresponding to the partial Hall constants $R_{Hi}\; (i=1,2)$ and a very sharp, asymmetrical, transition between them, situated just below $x=1/2.$
The transition becomes sharper and a width of the transition region decreases with an increase of $H.$

Very unusual pictures are obtained for the "random parquet" model (Fig.2b). As follows from the plots for the Hall constant $R_H$ a rapid decrease begins just at very small $x.$ The width of this region also decreases with an increase of $H.$  Again the deviations from strictly linear dependence of $\rho_{et}$ on $H$ are most essential in the region of rapid decrease.  In the remaining wide region of concentrations, including the point of equal concentrations $x=1/2,$  one has very slow changes till the boundary concentration $x=1.$ All pictures are strongly asymmetrical on $x.$ A correlation between the LLMR and a behaviour of $R_H$ in the RPM is more complicate because in this model both $\rho_{ed}$ and $\sigma_{e0},$ containing in $R(x,H)\approx \sigma_{e0}\rho_{ed},$ have broad tails at intermediate concentrations \ci{11,13}. A rapid decrease of $R(x,H)$ at small concentrations correlates with the LLMR effect in this region, while an almost constant behaviour of the MR at intermediate concentrations and a rapid decrease at $x \approx 1$ (see, for example, Fig.4d) are connected with a joint contribution of $\sigma_{e0}$ and $\rho_{ed}.$ A more obvious correlation 
exists between $\rho_{ed}$ and $\rho_{et}$ (or $\sigma_{ed}$ and $\sigma_{et}$)
 due to the circle equations (11) and (7) (see the section 4). 

One can see from Fig.2(a-c) the relative widths of the transition regions and values of $R_H$ for three
models. The most sharp changes  $R_H$ achieves for the "effective medium" model,
 the "random droplet" model has the intermediate behaviour, meanwhile the "random parquet" model has a widest region of relatively small values.
At equal phase concentrations $x=1/2$ all models reproduce the exact value (21,22). It is interesting that the exact value of $R_H$ at $x=1/2$ in all three models is significantly smaller than the larger $R_{H2}$ and is close to $R_{H1}$.

The analogous plots for $R_H$ (see Fig.3 (a)-(c)) and the magnetoresistance $R$ (Fig.4 (a)-(c)) are constructed for inhomogeneous systems with $z=1$ and $\eta = -1.$ This choice of parameters corresponds in our case to the restriction $\langle \mu \rangle =0$ and to homogeneous diagonal parts $\sigma_{10} = \sigma_{20}$. In this case the parameters of the MDT and the circles equations
have the next form
$$
\sigma_{1d} = \sigma_{2d}, \quad \sigma_{1t} = -\sigma_{2t}, \quad
z_0 = 0 =z_{\rho}, 
$$
$$
a=R= \sigma_{1t} \sqrt{1+H^2}= \fr{\sigma_{10} H}{\sqrt{1+H^2}}, \quad R_{\rho} = 1/R, \quad \sigma_{dm} = \sigma_{d0}.
\en(23)
$$
They also support a realization of non-monotonous behaviour of $\sigma_{ed}.$
The constructed plots show the existence of the Hall transition and a linear MR in this case too, but now a region of the SHT is situated symmetrically around the point $x = 0.5$ and the form of its $x$-dependencies and the amplitudes of the MR are, respectively, more smoother (especially for the RPM, which again deviates from the other models) and smaller than in the strongly inhomogeneous case $z/\eta \ll 1$. At $z\ne 1, \; \eta =-1$ a symmetry of the plots is lost. 

A comparison of the considered cases shows that, though the restriction $\langle \mu \rangle =0$ is enough for the appearance of the linear MR and the Hall transition, a necessary condition
for the existence of the pronounced LLMR and SHT is the strong inhomogeneity of the partial Hall conductivities $|\sigma_{1t}| \gg |\sigma_{2t}|$ or the partial Hall constants $z/\eta = R_{1H}/R_{2H} \ll 1$. In this case all formulas for the parameters of the MDT and the circles (7),(11) drastically simplify
$$
R \approx z_0 \approx \sigma_{1t}/2, \quad R - z_0 = a \approx -\sigma_{2t}, \quad R(1/2,H) \approx |\eta|H, 
$$
$$
R_{\rho} \approx |z_{\rho}|  \approx 1/2|\sigma_{2t}|, \quad R_H(1/2,H) \approx 1+ \eta.
\en(24)
$$
and one obtains for the exact values $\sigma_{dm}, \; \sigma_{tm}$ and $\rho_{dm}, \;\rho_{tm}$
$$
\sigma_{dm} \approx |\sigma_{tm}| \approx \sigma_{1t}/2 \approx 1/2H, 
$$
$$
\rho_{dm} \approx |\rho_{tm}| \approx |\rho_{2t}|/2 \approx 1/2|\sigma_{2t}| \approx |\eta| H/2z.
\en(25)
$$
Note that $\sigma_{dm} \gg \sigma_{ed}(1/2,H)$ and they coincide only at $z=1=-\eta.$ 
Analogously, one obtains for $\sigma_{d0}$ and $\rho_{d0}$
$$
\sigma_{d0} \approx \sqrt{\sigma_{1t} |\sigma_{2t}|}, \quad
\rho_{d0} \approx \sqrt{|\rho_{1t} \rho_{2t}|} \approx 1/\sigma_{d0}.
\en(26)
$$
These values agree with those from the corresponding plots from \ci{11,12} and from the Figures 2-4. Since $\sigma_{et}$ monotonically increases ($\rho_{et}$ decreases) with an increase of $x$  and $\sigma_{ed}$ ($\rho_{ed}$)
has only one maximum, one can determine $x_m$ (or $x_{m1}$) from the condition 
$\sigma_{ed} = \sigma_{et}$ (or $\rho_{ed} = \rho_{et}$).
 
\bs

\und{\bf 4. Correlations between LLMR and SHT}

\bs
The constructed plots unambiguously show an existence of strong correlations between the large magnetoresistance effect and a rapid, step-like, decrease of the Hall resistivity (or the Hall constant) in inhomogeneous two-phase media.
As it follows from the plots in \ci{11}, the analogous behaviour and correlation exist also for effective diagonal and Hall conductivities. These behaviour and correlations are a direct consequence of the exact circle relation (7). To show this let us consider $\sigma_{ed}$ and
$\sigma_{et}$ as functions of the concentration $x$ and differentiate (7) on $x$ twice. One obtains two relations
$$
\sigma_{ed} \sigma_{ed}' + (\sigma_{et} - z_0) \sigma_{et}'=0,
\en(27)
$$
$$
(\sigma_{et} - z_0) \sigma_{et}''- \sigma_{et}'^2 = \sigma_{ed} \sigma_{ed}'' + \sigma_{ed}'^2,
\en(28)
$$
where a prime denotes a differentiation on $x.$ From the first relation one can express $\sigma_{et}'$ through $\sigma_{ed}$ and $\sigma_{ed}'$
$$
\sigma_{et}' = \sigma_{ed} \sigma_{ed}'/(z_0 - \sigma_{et}) =
\sigma_{ed} \sigma_{ed}'/\sqrt{R^2 - \sigma_{ed}^2}
\en(29)
$$
It follows from (29) that the ratio $\sigma_{ed}'/(z_0 - \sigma_{et})$ on the left side is finite at the point $x_m,$ where $z_0 = \sigma_{et},$  since $\sigma_{ed}$ is maximal at this point and, consequently, $\sigma_{ed}' =0$ there also.
Then, from (27) one obtains that
$$
\sigma_{et}'^2(x_m) = -\sigma_{ed}(x_m) \sigma_{ed}''(x_m) = \sigma_{ed}(x_m) |\sigma_{ed}''(x_m)|,
\en(30)
$$
because $\sigma_{ed}''(x_m) \le 0.$ It follows from (30) that a derivative $\sigma_{et}'(x_m)$ is a geometric mean of $\sigma_{ed}(x_m)$ and $\sigma_{ed}''(x_m)$ and is proportional to $(\sigma_{dm})^{1/2}.$

As we already noted above, in the high $H$ limit $z_0$ is always situated between
$\sigma_{it},$ allowing a realization of the $\sigma_{ed}$ maximal
value. Moreover, though $\sigma_{it} \gg \sigma_{id}$ at high
magnetic fields, $\sigma_{dm}$ has the same order as
$\sigma_{it}.$ Thus, due to the exact circle equation (7), at the
high $H$ there is always a region of concentrations, where $\sigma_{ed}(x,H)$ takes its maximal value $\sigma_{dm} \approx \sigma_{1t}/2 \approx (2H)^{-1}.$ 
Just in this region
$\sigma_{et}'(x,H)$ becomes large giving a rapid change of $\sigma_{et}(x,H).$
A position and a width of this region depend on a model \ci{10,11}. Passing to 
$\hat \rho_e$ and the Hall constant one can check that the existence
of this region of concentrations is a necessary condition for an observation of the linear magnetoresistance, since $\rho_{ed}$ can have linear
behaviour at high magnetic fields only if $\sigma_{ed} \sim
H^{-1}$ together with $\sigma_{et}$ (for example, for $\sigma_{ed}
\sim H^{-2}$ the diagonal $\rho_{ed} \sim H^0$). 

The correlation between
large $\sigma_{ed}(x_m)$ and a rapid change of $\sigma_{et}(x,H)$ with a change of $x$ induces an analogous correlation between large $\rho_{ed}$ and a rapid change of the Hall resistivity $\rho_{et}(x,H),$
since they obey the similar circle equation (11) and the arc, where $\hat \rho_e$ takes its values, also covers almost the whole right semicircle. 
Such behaviour and correlations are very similar to those in the quantum Hall effect (QHE) \ci{17,18} with a difference, that in the QHE the jumps take place at low temperatures in
the dependencies of $\sigma_{xy}$ (or $\rho_{xy}$) on a magnetic field  and a height of the jumps (or steps) is determined by the quantum
conductance $\sigma_q,$ while in the SHT in two-phase systems such jump takes place up to room $T$ in the $x$-dependence of $R_H$ and a height of the jump is determined by the difference of the partial Hall constants $R_{H2} - R_{H1} = R_{H2}(1-z/\eta).$ 

A physical reason of this behaviour and strong correlations is the following. At high $H,$ in a homogeneous, badly conducting, medium the Hall currents are much larger than the Ohmic currents. When a concentration of the phase with a better conductivity increases and a medium becomes inhomogeneous, the Hall currents become {\it chaotic, strongly fluctuating,} and generate the Ohmic currents of the same order as they are ($\sim 1/H$) inside the good conducting phase. The most evident confirmation of this phenomenon one can see from (25) in the systems with opposite signs of Hall conductivities at the concentration $x_0,$ when $\sigma_{et}=0,$ but 
$\sigma_{ed} = \sqrt{R^2 - z_0^2} \approx \sqrt{\sigma_{1t}|\sigma_{2t}|} \ne 0.$  This gives a relatively large $\sigma_{ed} \sim 1/H$ and, consequently, a large linear magnetoresistance $R(x,H)\sim H.$ At the concentrations, when a good conducting phase forms rather large (bulk) regions, the Hall currents of this phase begin to prevail, thus reducing the magnetoresistance and the Hall resistivity. Than the larger difference between the conducting properties of two phases the larger and sharper becomes a jump of the Hall constant and the larger
magnetoresistance appears in these binary systems.

If one neglects $\sigma_{ed}$ in such inhomogeneous systems (with $\sigma_{it} \gg \sigma_{id}$), then it follows from (7),(19) that only the partial Hall currents can exist in this system with an abrupt jump between them at some undetermined concentration (usually one believes that it takes place at the percolation edge $x=1/2$). 
Our results show that just $\sigma_{ed} \ne 0$ is responsible for a more smooth and realistic transition between two Hall plateaus in these systems, though it can be very sharp in the strongly inhomogeneous case. The three considered model demonstrate that the structure of random inhomogeneities plays also an essential role in a form and a position of this transition on the concentration axis.

\bs

\und{\bf 5. Conclusion}
\bs

Using three explicit approximate expressions for the effective
conductivity of 2D isotropic two-phase systems in a magnetic
field, we have studied properties of the effective resistivities $\rho_{ed,et}(x,H)$ and the Hall constant $R_H(x,H)$
of planar inhomogeneous isotropic two-phase
systems. The  plots of the $x$- and $H$-dependencies of
$\rho_{et}(x,H)$  and $R_H(x,H)$
at some values of the inhomogeneity parameters 
$z$ and $\eta$  are constructed. For strongly inhomogeneous (already at $z \approx 10^{-2}, \; \eta =1$) the "random droplets" and effective medium models
show  a sharp transition between two
Hall plateaus at intermediate concentrations ($0.2 \le x \le 0.4$ for RDM and $0.4 \le x \le 0.5$ for EMM), while in the "random parquet" model this transition begins at very small concentrations.
At $\eta =-1$ and $z=1$ the more smoother Hall transition takes place in the region of concentrations situated symmetrically around $x=1/2.$
This behaviour of the constructed plots strongly correlates with the LLMR effect,
which takes place in these models in the same regions of concentrations, where a sharp transition exists. 
Both these effects follow from the exact duality and take place at $\langle \mu \rangle \ne 0$ as well as at $\langle \mu \rangle =0,$ but in the case $z \ll 1$
they are more pronounced. These properties are qualitatively compatible with the known experimental data (see, for example \ci{19}).
A possible physical explanation of such behaviour is proposed.
We hope that the obtained results, based on the exact dual symmetries, can be applied also for a description of magneto-transport properties of various real inhomogeneous systems (regular and non-regular as well as random),
satisfying the symmetries and having the similar structures as the considered models, in a wide range of inhomogeneity parameters and at arbitrary concentrations and magnetic fields.

\bs
\und{ Acknowledgments}

\bs
The author is thankful to Prof.F.V.Kusmartsev for useful discussions and a warm hospitality at the Loughborough University.
The work was supported by the RFBR grant 02-02-16403, by the ESF network AQDJJ and by the EPSRC grant GR/S05052/01.

\end{document}